\documentclass[conf]{new-aiaa}
\usepackage[utf8]{inputenc}

\usepackage{graphicx}
\graphicspath{{figs/}}
\usepackage{dcolumn}
\usepackage{bm}
\usepackage{subfigure}

\usepackage{algorithm}
\usepackage{algorithmic}

\usepackage{soul}
\usepackage{amsmath,amsfonts}
\usepackage{graphics}
\usepackage{color}
\usepackage{xcolor}
\definecolor{b}{rgb}{0,0,1}
    
\usepackage{enumitem}

\usepackage{comment}
\usepackage{hyperref}
\usepackage{lipsum}
\usepackage{tabularx}
\usepackage{stackengine}

\title{Hyperparameter Search using Genetic Algorithm for Surrogate Modeling of Geophysical Flows}

\author[1]{Suraj Pawar \footnote{Ph.D. Candidate, School of Mechanical \& Aerospace Engineering, Oklahoma State University, Stillwater, OK 74078, USA.}}
\author[1]{Omer San \footnote{Associate Professor, School of Mechanical \& Aerospace Engineering, Oklahoma State University, Stillwater, OK 74078, USA.}}
\author[2]{Gary G. Yen \footnote{Regents Professor, School of Electrical \& Computer
Engineering, Oklahoma State University, Stillwater, OK 74078, USA.}}
\affil[1]{School of Mechanical \& Aerospace Engineering, Oklahoma State University, Stillwater, OK 74078, USA.}
\affil[2]{School of Electrical \& Computer Engineering, Oklahoma State University, Stillwater, OK 74078, USA.}

\begin{document}

\maketitle

\begin{abstract}
The computational models for geophysical flows are computationally very expensive to employ in multi-query tasks such as data assimilation, uncertainty quantification, and hence surrogate models sought to alleviate the computational burden associated with these full order models. Researchers have started applying machine learning algorithms, particularly neural networks, to build data-driven surrogate models for geophysical flows. The performance of the neural network highly relies upon its architecture design and selection of other hyperparameters. These neural networks are usually manually designed through trial and error to maximize their performance. This often requires domain knowledge of the neural network as well as the problems of interest. This limitation can be addressed by using an evolutionary algorithm to automatically design architecture and select optimal hyperparameters of the neural network. In this paper, we apply the genetic algorithm to effectively design the long short-term memory (LSTM) neural network to build the non-intrusive reduced order model of the sea-surface temperature field.
\end{abstract}

\section{Introduction}
There are numerous tasks in engineering design, control, and climate modeling that require solving a partial differential equation (PDE) based forward model for many ensembles. For example, in sequential data assimilation, an ensemble of models is run to approximate the covariance matrix \cite{lewis2006dynamic}. As the desired level of accuracy increases, the resolution of the spatiotemporal numerical discretization also increases. This can cause a significant increase in the computational resources and can become the bottleneck in the outer loop of the design and forecast cycle. For example in computational fluid dynamics (CFD), the three-dimensional simulation of turbulent flows \cite{moin1998direct} is often used to obtain an exact solution to the Navier-Stokes equation but is less frequently used in related tasks such as shape optimization \cite{polat2013aerodynamic}. Therefore, there is a sustained interest in developing a reduced-order model (ROM) that is computationally much cheaper than the full order model (FOM) and also provides the solution with a sufficient level of accuracy \cite{lucia2004reduced,taira2017modal,benner2015survey}.

In recent years, there is a growing interest in developing non-intrusive ROMs for physical systems \cite{xiao2015non,hesthaven2018non,pawar2019deep}. The non-intrusive refers to the use of only data to construct the ROM. The non-intrusive ROMs are particularly attractive for systems where the perfect model is unknown, but there is an abundance of data available for that system. This situation is commonplace for geophysical flows. The geophysical system model can be imperfect due to a variety of reasons such as approximation of subgrid-scale processes due to insufficient grid resolution, uncertainty in model parameters or incorrect structure of the model itself \cite{schneider2017earth}. However, there is an abundance of data from remote sensing, satellite observations, and experimental measurements for geophysical flows. As a result, there have been several works to harness this data for the effective forecast of geophysical processes \cite{o2018using,reichstein2019deep}. 
 
In this work, we focus on the use of linear dimensionality reduction technique followed by the machine learning algorithm to evolve the latent space \cite{mohan2018deep,rahman2019nonintrusive,pawar2020data}. Specifically, we use proper orthogonal decomposition (POD) to identify the latent space of the FOM, and then the long short-term memory (LSTM) neural network is utilized to model the evolution of latent space. One of the main advantages of this surrogate modeling technique is that it is purely data-driven and hence is particularly appealing to geophysical flows. The archival data collected from remote sensing and in-situ measurements can be exploited to build the surrogate model and then the surrogate model can be employed for the forecast task or in the data assimilation cycle. As the new data becomes available, the surrogate model can be retrained using the transfer learning approach \cite{tan2018survey} to improve the forecast performance. 

Apart from ROM, the use of the deep neural network has significantly increased for many scientific applications \cite{mjolsness2001machine,brunton2020machine,pawar2021data}. One of the main challenges with neural networks is that their performance is highly dependent on their architecture \cite{krizhevsky2017imagenet}. Additionally, the neural network has a large number of hyperparameters that are problem-dependent. Usually, the neural networks are manually designed through trial and error and this procedure can be time-consuming. Moreover, rich domain knowledge about the data and neural networks is required to achieve a promising performance. There are methods like grid search \cite{bergstra2011algorithms} or a random search \cite{bergstra2012random} to find a good combination of hyperparameters. But these methods are not scalable as the dimension of the search space goes up. Hence, there is a growing interest in automating the design of deep neural network architecture and hyperparameter search that will allow the user with no specialized knowledge of the neural network to apply to their problem of interest \cite{elsken2018neural}. In the present study, we use the genetic algorithm (GA) to optimize the architecture design and hyperparameters of the LSTM neural network. Readers are referred to a recent survey by \citeauthor{liu2021survey} \cite{liu2021survey} on most recent evolutionary computation-based neural architecture search methods in light of the core components, to systematically discuss their design principles and justifications on the design. The LSTM network is used within the surrogate model design of the NOAA sear-surface temperature (SST) data set.  

The rest of the paper is structured as follows. Firstly, we describe the non-intrusive ROM methodology and data pre-processing in Section~\ref{sec:nirom}. Then, the details of the proposed algorithm are provided in Section~\ref{sec:ga}. Section~\ref{sec:experiments} discusses the experimental results and their analysis. Finally, in Section~\ref{sec:conclusion}, conclusions and future works are outlined.

\section{Non-intrusive Reduced Order Modeling} \label{sec:nirom}

\subsection{Proper orthogonal decomposition}
We use the proper orthogonal decomposition (POD) to extract the dominant modes from the data. We collect the data snapshots $\mathbf{u}_1, \mathbf{u}_2, \dots, \mathbf{u}_N$ $\in$ $\mathbb{R}^M$ at different time instances. Here, $M$ is the spatial degree of freedom which is equal to the total number of grid points, and $N$ is the number of snapshots. In POD, we construct a set of orthonormal basis functions that optimally describes the field variable of the system in $L_2$-norm. The snapshot data matrix $\mathbf{A}$  is formed as
\begin{equation}
    \mathbf{A} = [\mathbf{\tilde{u}}_1 | \mathbf{\tilde{u}}_2 | \dots | \mathbf{\tilde{u}}_N ]  \in \mathbb{R}^{M \times N},
\end{equation}
where the mean-subtracted (or anomaly) fields  $\mathbf{\tilde{u}}_i \in$ $\mathbb{R}^M$ are computed as follows 
\begin{equation}
    \mathbf{\tilde{u}}_i = \mathbf{u}_i - \bar{\mathbf{u}}, \quad \bar{\mathbf{u}} = \frac{1}{N} \sum_{i=1}^N \mathbf{u}_i,
\end{equation}
where $\bar{\mathbf{u}}$ is the time-average of the solution field. Once the snapshot data matrix is formed, we use singular value decomposition (SVD) to compute the left and right singular vectors of the matrix $\mathbf{A}$. In matrix form, the SVD can be written as 
\begin{equation}
    \mathbf{A} = \mathbf{W}\mathbf{\Sigma}\mathbf{V}^T = \sum_{k=1}^N\sigma_k \mathbf{w}_k \mathbf{v}_k^T,
\end{equation}
where $\mathbf{W}\in \mathbb{R}^{M \times N}$, $\mathbf{\Sigma} \in \mathbb{R}^{N \times N}$, and $\mathbf{V} \in \mathbb{R}^{N \times N}$. The $\mathbf{W}$ and $\mathbf{V}$ contains the left and right singular vectors which are identical to the eigenvectors of $\mathbf{A} \mathbf{A}^T$ and $\mathbf{A}^T \mathbf{A}$, respectively. Also, the square of singular vales are equal to the eigenvalues, i.e., $\lambda_k = \sigma_k^2$. The vectors $\mathbf{w_k}$ (also the eigenvectors of $\mathbf{A} \mathbf{A}^T$) are the POD basis functions and we denote them as $\phi_k$ in this text. The POD basis functions are orthonormal (i.e., $\langle \phi_i, \phi_j \rangle = \delta_{ij}$, where $\delta_{ij}$ is the Kronecker delta) and are computed in an optimal manner in the $L_2$ sense \cite{holmes2012turbulence, rowley2017model}. The state of the dynamical system can be approximated using these POD basis functions as follows
\begin{equation} \label{eq:rom_construction}
    \mathbf{u}(\mathbf{x}, t) = \bar{\mathbf{u}} + \sum_{k=1}^{R} a_k(t) \phi_k(\mathbf{x}),
\end{equation}
where $R$ is the number of retained basis functions such that $R << N$ and $a_k$ are the time-dependent modal coefficients. The POD basis functions minimize the mean-square error between the field variable and its truncated representation. The number of retained modes is usually decided based on their energy content. Using these retained modes, we can form the POD basis set $\mathbf{\Phi}=\{ \phi_k\}_{k=1}^{R}$ to build the ROM.

\subsection{Model-free evolution of the ROM}
The model-free prediction of the evolution of ROMs has a promising application, especially for geophysical dynamical systems. For a variety of tasks such as data assimilation, uncertainty quantification of geophysical flows, the ensemble of the forward model needs to be run and this can be computationally expensive. Instead of evolving the forward model, the surrogate model based on ROMs can be utilized. Also for many geophysical dynamical systems, the mechanistic description of the dynamical system is unavailable or incomplete due to coarse grid resolution. However, there is an abundance of data obtained from local and satellite observations for many past decades. In recent years, data-driven approaches exploiting the recurrent neural network (RNN) are proven successful for the model-free prediction of spatiotemporal chaotic dynamical systems \cite{pathak2018model,vlachas2018data}. 

The time-dependent modal-coefficients given in Equation~\ref{eq:rom_construction} are obtained by projecting the mean subtracted field on the POD basis as follows 
\begin{equation}
    a_k(t_i) = \langle \mathbf{\tilde{u}}_i ;\phi_k\rangle,
\end{equation}
where angle parenthesis denotes the inner product of two functions. Therefore, if we can model the prediction of the evolution of modal coefficients we can reconstruct the solution field in the future. One of the popular approaches of time-series prediction is the RNN and has been recently applied to several studies on non-intrusive ROMs for physical systems \cite{mohan2018deep,maulik2020time,ahmed2019memory,wan2018data}. This is motivated by the fact that RNN takes into account the past history of the system for future state prediction and the availability of historical data to train the neural network. The long short-term memory (LSTM) neural network is one of the most successful variants of the RNNs that solves the vanishing gradient problem \cite{hochreiter1997long}. In this study, we consider an input sequential data matrix $\mathcal{X}$ and the output sequential data matrix $\mathcal{Y}$. Each sample of the input training matrix $\mathcal{X}$, i.e., $\mathcal{X}_n$ is constructed as $\left\{a_{1}^{(k)} ,\dots, a_{R}^{(k)}; \ \dots \ ; a_{1}^{(k-\sigma+1)},\dots, a_{R}^{(k-\sigma+1)}\right\}$ and the corresponding output sample in output sequential data matrix $\mathcal{Y}$, i.e., $\mathcal{Y}_n$ is $\left\{a_{1}^{(k+1)} ,\dots, a_{R}^{(k+1)}\right\}$, where $k$ corresponds to the time index, and $n$ is for the $n$th sample. The parameter $\sigma$ is called the lookback time-windows. Since we are using the information of only $\sigma$ past temporally consecutive states as the input, the LSTM can capture dependencies up to $\sigma$ previous time steps.  


\subsection{Data preparation}
Model-free prediction is particularly important for real-world data which is a manifestation of many multiscale processes that cannot be approximated accurately with any of the model-based approaches. The dynamics of the flow in real-life is governed by the coupling with external systems, and this coupling might be unknown or cannot be modeled. Therefore, the real-world data is not well-behaved and non-intrusive methods are well suited for these complex flows. To this end, we investigate the application of non-intrusive ROMs for the NOAA Optimum Interpolation Sea Surface Temperature (SST) V2 data set\footnote {\href{https://psl.noaa.gov/}{https://psl.noaa.gov/ (Accessed 2020)}}. This data set consists of the weekly average sea surface temperature (SST) data snapshots on a one-degree resolution grid and is produced by combining satellite and local measurements. The seasonal fluctuations give rise to a strong periodic structure to the temperature field in this data set.  

Each snapshot of the data set has the dimension of $180 \times 360$ along the latitudes and longitudes with one-degree grid resolutions. The data points corresponding to land are removed by using the mask operation and the surrogate model is built for the flattened data corresponding to only the ocean surface. This data is available from October 22, 1981, to June 
30, 2018 (i.e., total 1,914 snapshots). 

\begin{figure}[htbp]
\centering
\includegraphics[width=0.45\textwidth]{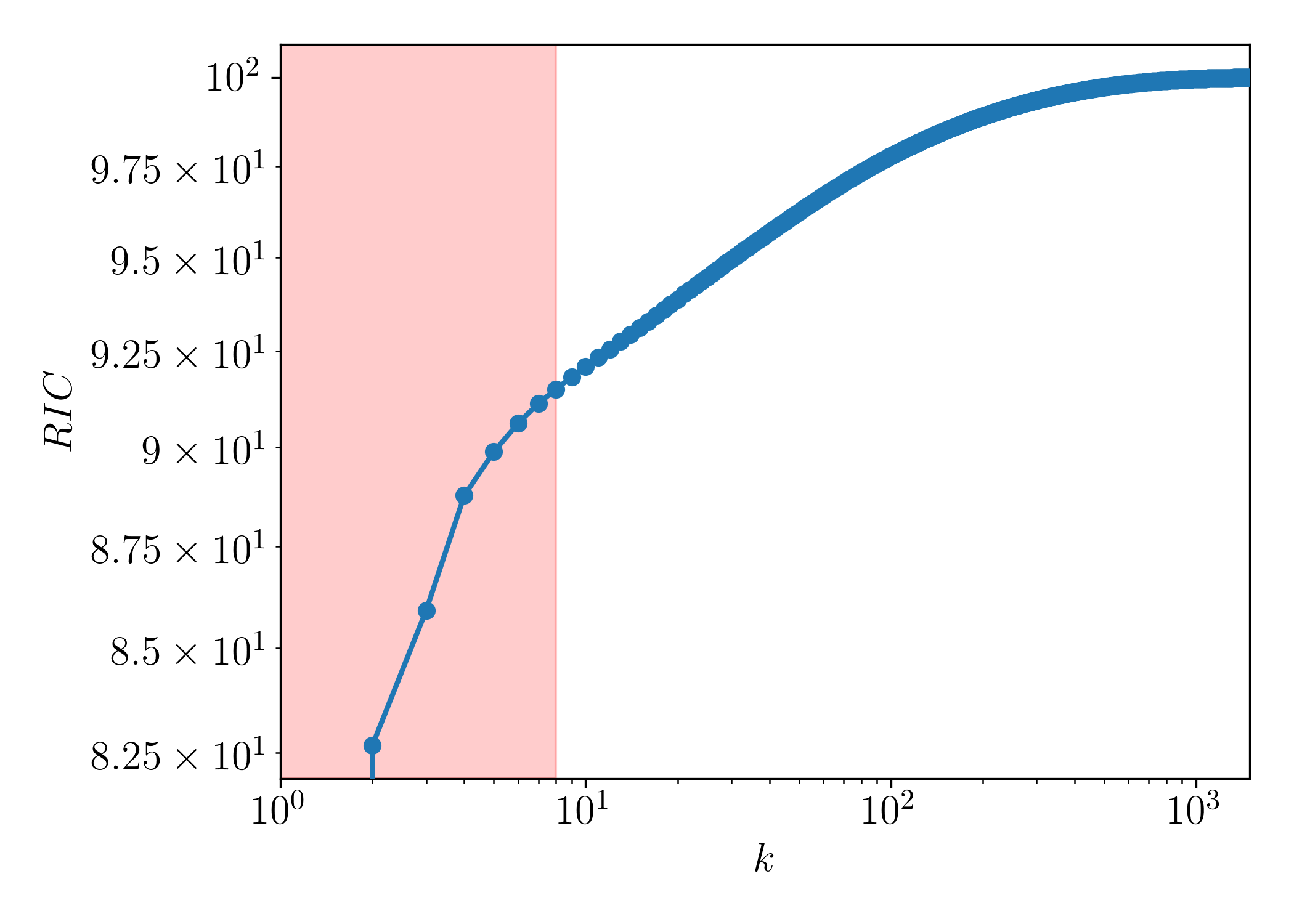}
\caption{Percentage of the square of singular values of the snapshot data matrix $\mathbf{A}$ (equivalent to eigenvalues of $\mathbf{A}\mathbf{A}^T$ or $\mathbf{A}^T\mathbf{A}$). The number of retained modes for the ROM is $R=8$.}
\label{fig:ric}
\end{figure}

Once the data prepossessing is done, then the snapshot data matrix $\mathbf{A}$ is formed using the mean-subtracted anomalies of the temperature field. The number of modes for the ROM, R, is selected based on a relative information content (RIC) formula as given below
\begin{equation}
    \text{RIC}(R) = \left(\frac{\sum_{k=1}^{R}\sigma_k^2} {\sum_{k=1}^{N}\sigma_k^2}\right).
\end{equation}
The RIC represents the fraction of information (variance) of the total data that can be recovered using $R$ basis functions. Figure~\ref{fig:ric} shows the RIC percentage for the NOAA SST data set. We fix the number of retained modes to be $R=8$ that captures around 92\% (i.e., $9.182 \times 10^1$) of the variance of the data and these modes are sufficient to capture the seasonal trends in the SST data set. It can also be seen that the increase in the number of modes after 8 modes gives a very small increase in the total information variance, as these modes are mainly responsible for capturing small-scale fluctuations. 

\section{Genetic Algorithm for LSTM Network} \label{sec:ga}
\begin{figure}[htbp]
\centering
\includegraphics[width=0.75\textwidth]{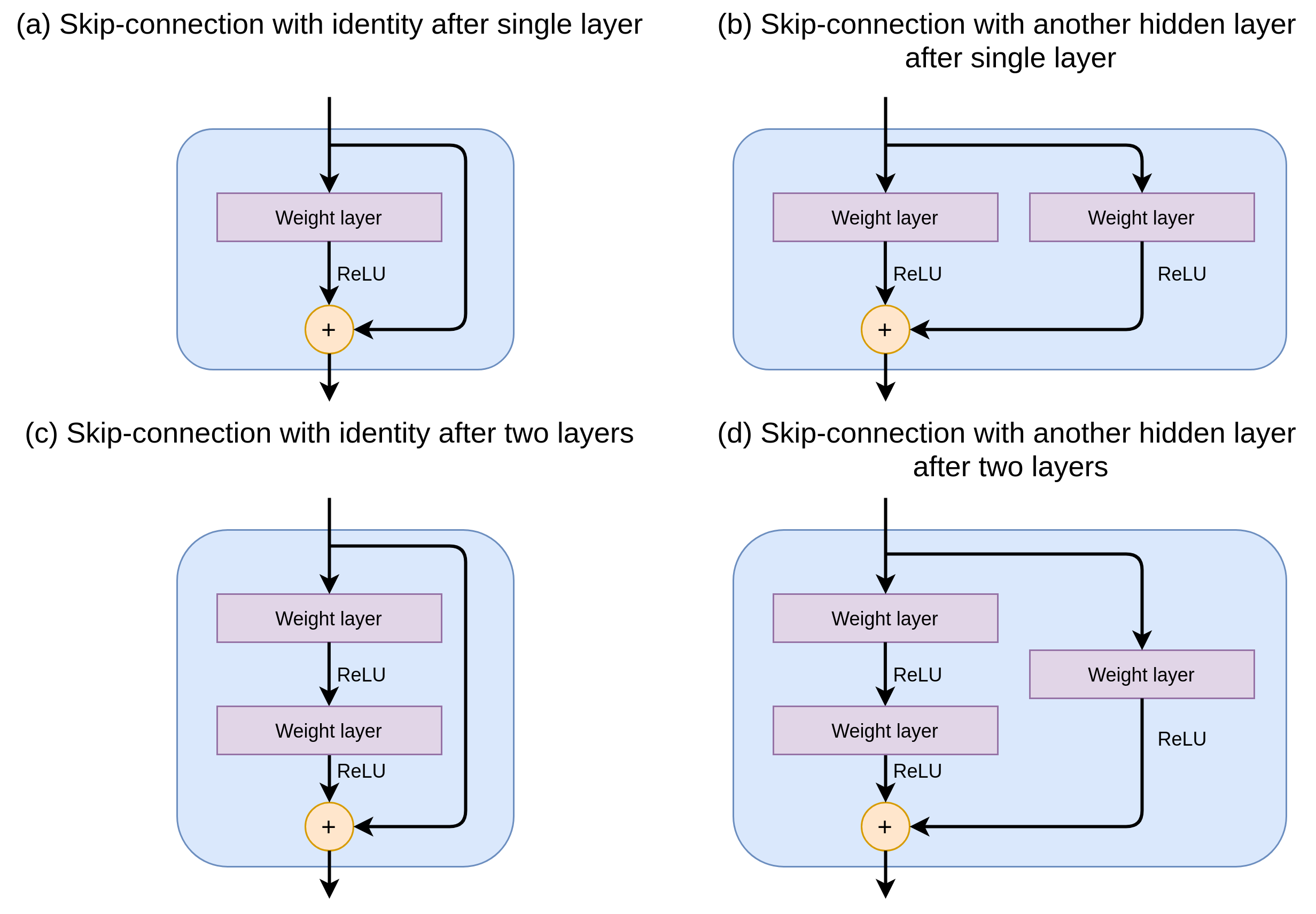}
\caption{Different types of encoded small architectures that act as the building block of the entire LSTM architecture.}
\label{fig:lstm_blocks}
\end{figure}

The performance of the neural network is highly dependent on the design of the neural network architecture and the selection of other hyperparameters such as activation function, optimizer, weight initialization, etc. Similar to other hyperparameter optimization studies \cite{xie2017genetic,sun2020automatically}, we encode the entire LSTM architecture using different types of small architectures. Figure~\ref{fig:lstm_blocks} shows different types of small architectures that act as the building blocks of the main LSTM network. In each of these small architectures, skip-connection is employed as it allows training a deep neural network without the vanishing gradient problem \cite{drozdzal2016importance,he2016deep}. The final LSTM network is constructed by ordering these building blocks. The number of LSTM cells and the number of building blocks are the two additional parameters related to the design of the neural network architectures that need to be optimized using the genetic algorithm. The other hyperparameters to optimize are the type of the optimizer (for example, SGD and Adam), the learning rate of the optimizer, type of initialization for weights and biases such as RandomUniform and Xavier initialization, and the type of activation function (for example, ReLU and tanh). 

Algorithm~\ref{alg:ga} lists the different steps of the proposed framework. We start by initializing the population in a given size. The LSTM network is randomly constructed for each individual from the predefined building blocks and other hyperparameters are assigned based on the assignment. The neural network is evaluated using the three-fold cross-validation on the training dataset. Once the fitness (i.e., the validation mean squared error in this study) is evaluated for each individual in the population, they are sorted based on their fitness values. From this population, a certain percentage of elite individuals (i.e., the individuals with better fitness values) are retained and some individuals are selected using the tournament selection for offspring generation. The offspring are generated from the selected parents using crossover and mutation operators. The new population is formed by combining the retained elite individuals from the old population and the newly generated offspring. The counter for the number of generations is increased by one and this procedure is continued for the specified number of generations.

\begin{algorithm}
 \caption{The proposed genetic algorithm }
 \begin{algorithmic}[1]\label{alg:ga}
 \renewcommand{\algorithmicrequire}{\textbf{Input:}}
 \renewcommand{\algorithmicensure}{\textbf{Output:}}
 \REQUIRE A set of predefined building blocks, population size, maximum number of generations, dataset for training, elite percentage to retain.
 \ENSURE  Discovered LSTM architecture along with other hyperparameters 
  \STATE $P_0 \leftarrow$ Initialize the population with the given population size. 
 \STATE $k \leftarrow 0$
  \FOR {$k = 0$ to $the~maximum~generation~number$}
  \STATE Evaluate the fitness of each individual within the population $P_k$;
  \STATE $P_k^E \leftarrow$ Retain the elite population with the maximum value of the fitness;
  \STATE $P_k^T \leftarrow$ Select the population for offspring generation using tournament selection;
  \STATE $Q_k \leftarrow$ Generate offspring from the selected population using the proposed crossover and mutation operations
  \STATE $P_{k+1} \leftarrow$ $P_k^E \cup  Q_k$
  \STATE $k = k + 1$
  \ENDFOR
 \RETURN the individual with the best fitness form $P_k$ 
 \end{algorithmic} 
 \end{algorithm}
 
Each individual in the population is defined using 6 integer numbers and 1 real number. These numbers represent the hyperparameter of the LSTM network. The upper and lower limit of these hyperparameters and the type of distribution used for generating these numbers are reported in Table~\ref{tab:hp}. The first parameter is the type of the building block of the network and four different building blocks as shown in Figure~\ref{fig:lstm_blocks} that can be used to design the LSTM network. The second and third parameters are related to how deep the LSTM network will be and how wide the neural network within the LSTM cell will be. The other parameters such as activation function, optimizer, learning rate, and weight initialization are related to the training of the LSTM network. During the evaluation, the LSTM network corresponding to each individual is trained and the validation mean squared error (MSE) is assigned as the fitness for that individual. The population is sorted based on the value of the fitness in ascending order (lower MSE means better fitness). 

\begin{table}[htbp]
\caption{Hyperaprameters of the LSTM network that are optimized}
\label{tab:hp}
\begin{center}
\begin{tabular}{|m{0.3\textwidth}|m{0.1\textwidth}|m{0.1\textwidth}|m{0.1\textwidth}|m{0.3\textwidth}|}
\hline
\textbf{Hyperparameter} & \textbf{\textit{Lower bound}}& \textbf{\textit{Upper bound}}& \textbf{\textit{Type}} & \textbf{\textit{Available options}}\\ 
\hline \vspace{1mm}
Type of building block &  1 & 4 & Uniform & Shown in Figure~\ref{fig:lstm_blocks}\\
\hline \vspace{1mm}
Number of hidden layers &  2 & 8 & Uniform & 2,3,4,5,6,7,8\\
\hline \vspace{1mm}
Number of LSTM cell in each block &  40 & 160 & Normal & Integer between 40 and 160\\
\hline \vspace{1mm}
Optimizer &  1 & 3 & Uniform & Adam, RMSprop, SGD\\
\hline \vspace{1mm}
Activation function &  1 & 3 & Uniform & ReLU, tanh, LeakyReLU\\
\hline \vspace{1mm}
Weight initialization &  1 & 3 & Uniform & RandomNormal, RandomUniform, GlorotNormal\\
\hline \vspace{1mm}
Learning rate of the optimizer &  $10^{-4}$ & $10^{-2}$ & Lognormal & Real number between $10^{-4}$ to $10^{-2}$\\
\hline
\end{tabular}
\label{tab1}
\end{center}
\end{table}

The next step is the selection of parents for the offspring generation. We use tournament selection to select parents. Algorithm~\ref{alg:selection} shows the procedure of tournament selection. In tournament selection, a few individuals are selected randomly from the population and the best individual is selected for offspring generation. 
\begin{algorithm}
 \caption{Parents selection}
 \begin{algorithmic}[1]\label{alg:selection}
 \renewcommand{\algorithmicrequire}{\textbf{Input:}}
 \renewcommand{\algorithmicensure}{\textbf{Output:}}
 \REQUIRE The population $P_k$, population size $N$, number of elite individuals $M$, tournament size $T$
 \ENSURE  Population for offspring generations $P_k^T$
  \STATE $P_k^T \leftarrow \emptyset $
  \FOR {$i = 0$ to $N-M$}
  \STATE $\mathbf{T} \leftarrow$ Randomly generate $T$ numbers without repetition;
  \STATE $P_k^T \leftarrow P_k^T \cup P_k[\text{min}(\mathbf{T})]$;
  \ENDFOR
 \RETURN The selected population $P_k^T$ 
 \end{algorithmic} 
 \end{algorithm}
 
The details of generating the offspring are provided in Algorithm~\ref{alg:crossover_mutation}. The offspring population is generated in two stages. The first stage is the crossover operation (lines 2 -13), and we utilize the uniform crossover operator. In the uniform crossover operation, two individuals are selected from the population. Then for every gene of the individual, a random number is generated, and if this number is greater than the crossover probability then the corresponding gene at the position $j$ of two individuals is swapped. During the mutation, the gene at the position $i$ is changed if the random number is greater than the mutation probability.  
\begin{algorithm}
 \caption{Offspring generation}
 \begin{algorithmic}[1]\label{alg:crossover_mutation}
 \renewcommand{\algorithmicrequire}{\textbf{Input:}}
 \renewcommand{\algorithmicensure}{\textbf{Output:}}
 \REQUIRE The selected population $P_k^T$ containing individual and their fitness, the probability of crossover operation $p_c$, mutation operation probability $p_m$.
 \ENSURE  The offspring population $Q_k$ 
 \STATE $Q_k \leftarrow \emptyset $
 \WHILE{$i < |P_k^T|$ }
 \STATE $p_1,p_2 \leftarrow$ Select two consecutive individuals from $P_k$ (i.e, $2i,2i+1$) 
 \FOR{$j$ = $0$ to $|p_1|$}
 \STATE $r \leftarrow$ Randomly generate a number between $[0,1]$;
 \IF{$r > p_c$}
 \STATE Swap the genes of $p_1$ and $p_2$ 
 \ENDIF
 \STATE $j \leftarrow j+1$
 \ENDFOR
 \STATE $Q_k \leftarrow Q_k \cup p_1 \cup p_2$;
 \STATE $i \leftarrow i+1$
 \ENDWHILE
 \FOR{each individual $p$ in $Q_k$}
 \STATE $r \leftarrow$ Randomly generate a number between $[0,1]$;
 \IF {$r < p_m$}
 \STATE $i \leftarrow$ Randomly choose a point in $p$;
 \STATE Change the gene at the point $i$ of $p$;
 \ENDIF
 \ENDFOR
 \RETURN Offspring population $Q_k$ 
 \end{algorithmic} 
 \end{algorithm}

\section{Numerical Experiments} \label{sec:experiments}
\begin{figure}[htbp]
\centering
\includegraphics[width=0.8\textwidth]{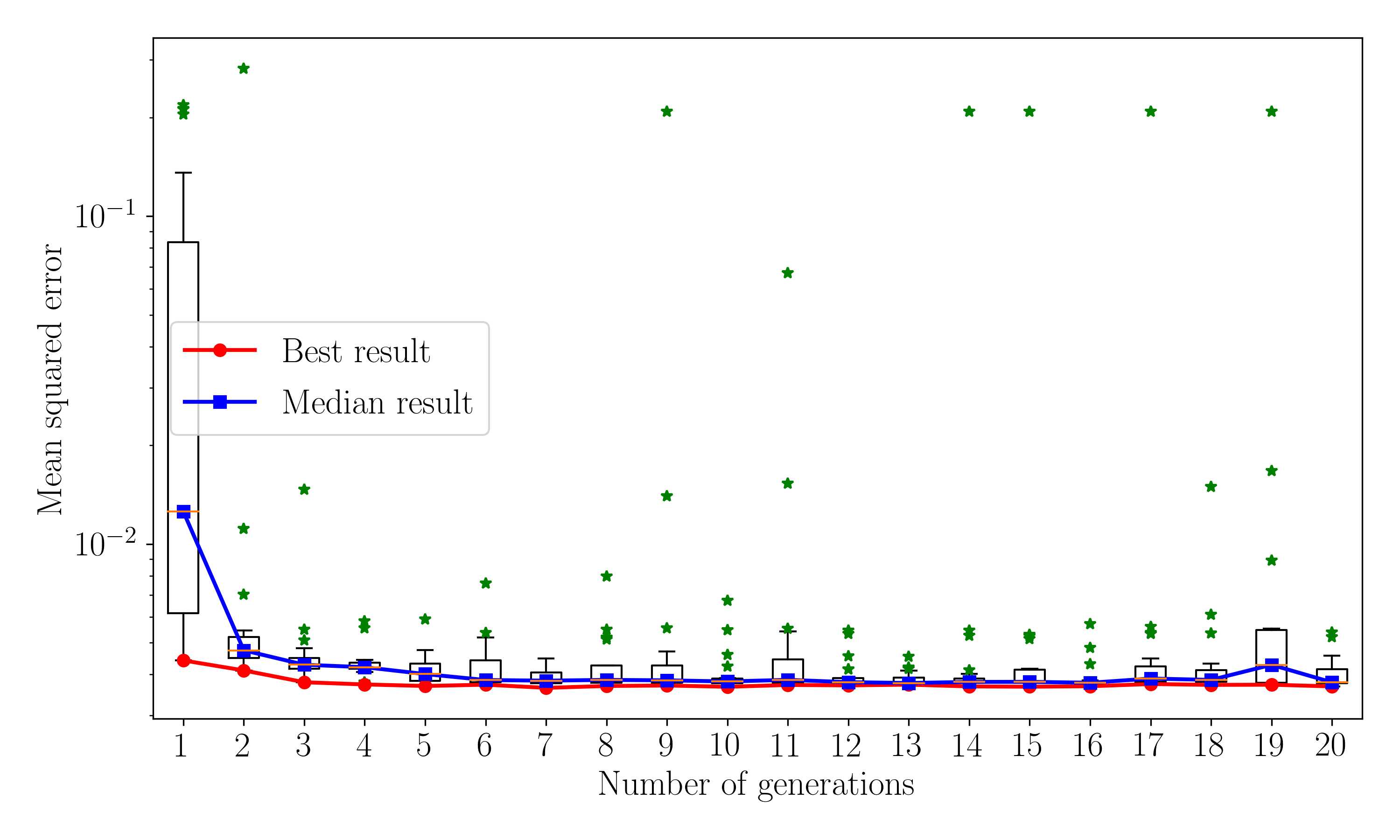}
\caption{The evolutionary trajectory of the proposed algorithm in discovering
the best architecture of the LSTM on the NOAA SST data set.}
\label{fig:boxplot}
\end{figure}

In this Section, we provide the result of the GA in optimizing the architecture design and hyperparameters of the neural network. Then, we show the performance of the optimized LSTM network in emulating and forecasting the SST. The NOAA SST data set is available from October 22, 1981, to June 30, 2018, which corresponds to 1,914 snapshots. We utilize 70\% of the data selected randomly from the first 1,500 snapshots as the training data set. During the hyperparameter optimization of the LSTM network, the individual within the population is trained for 100 epochs and three-fold cross-validation is used to avoid overfitting. The lookback time window of the LSTM network is set at $\sigma = 8$, which is found to capture the temporal correlation in modal coefficients. The fitness of each individual is the mean of the mean squared error (MSE) for the held-out data set from three folds. The batch size is fixed at 64 for the training. The population size for this experiment over the number of generations is set at 20. Figure~\ref{fig:boxplot} displays the evolution trajectory of the fitness of individuals within the population with generations of the GA. The boxplot is utilized to show the statistics of the population at each generation. Figure~\ref{fig:boxplot} also reports the median and minimum MSE of the population at each generation. As the evolution progresses, the MSE on the validation data set decreases. The height of the box indicates the variance of the fitness of the population at each generation, and we can observe that the variance is substantially low after the first generation. The MSE decreases sharply from the first generation to the second generation, which can be attributed to the random initialization of the population at the beginning of the GA. For the problem investigated in this study, the five to ten generations seem enough to find the best set of hyperparameters.     

\begin{figure}[htbp]
\centering
\includegraphics[width=0.75\textwidth]{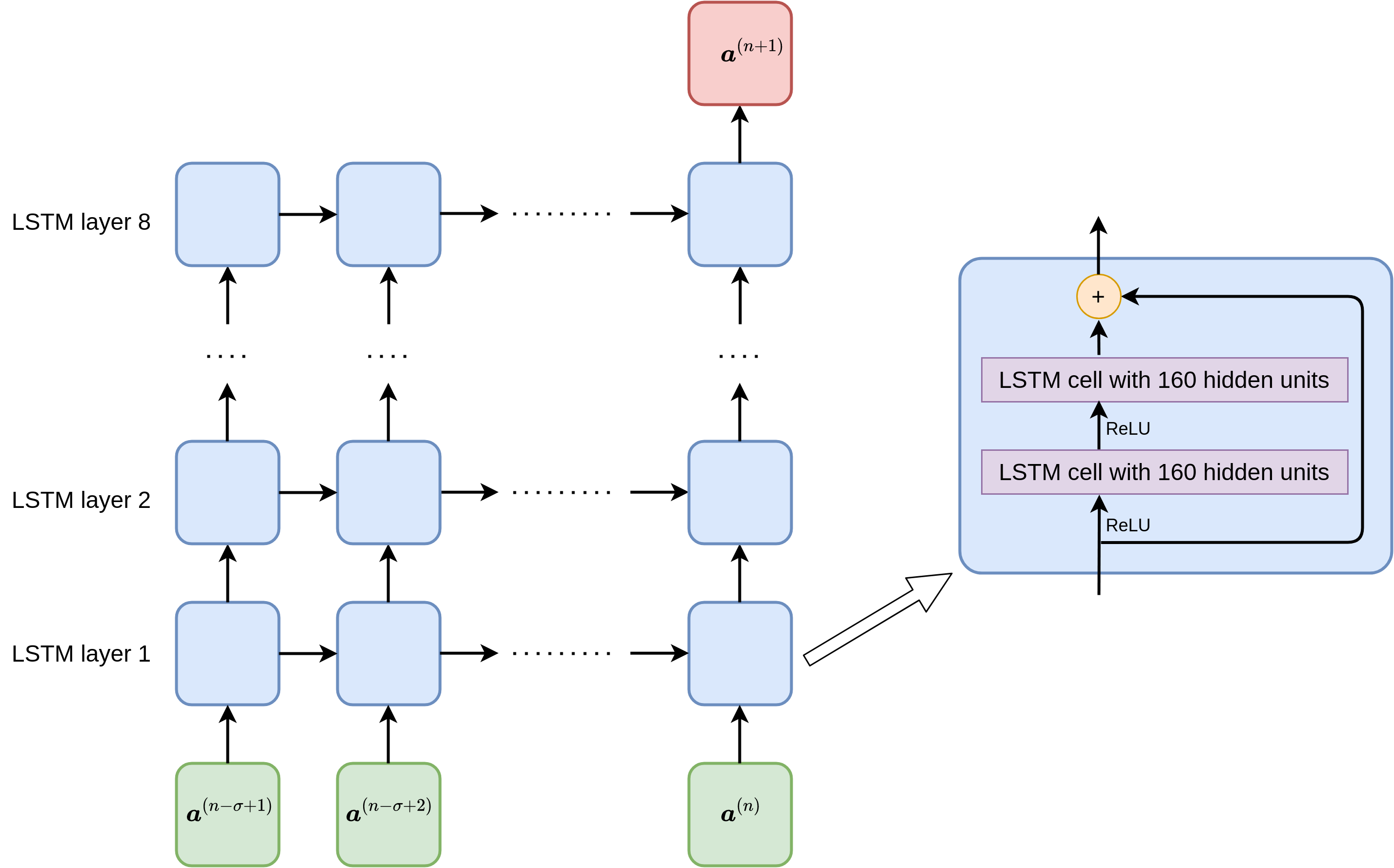}
\caption{Optimal neural network architecture discovered using the genetic algorithm. The selected hyperparameters are RandomNormal initialization, Adam optimizer with the learning rate = 0.0047. }
\label{fig:opt_architecture}
\end{figure}

\begin{figure}[htbp]
\centering
\includegraphics[width=0.9\textwidth]{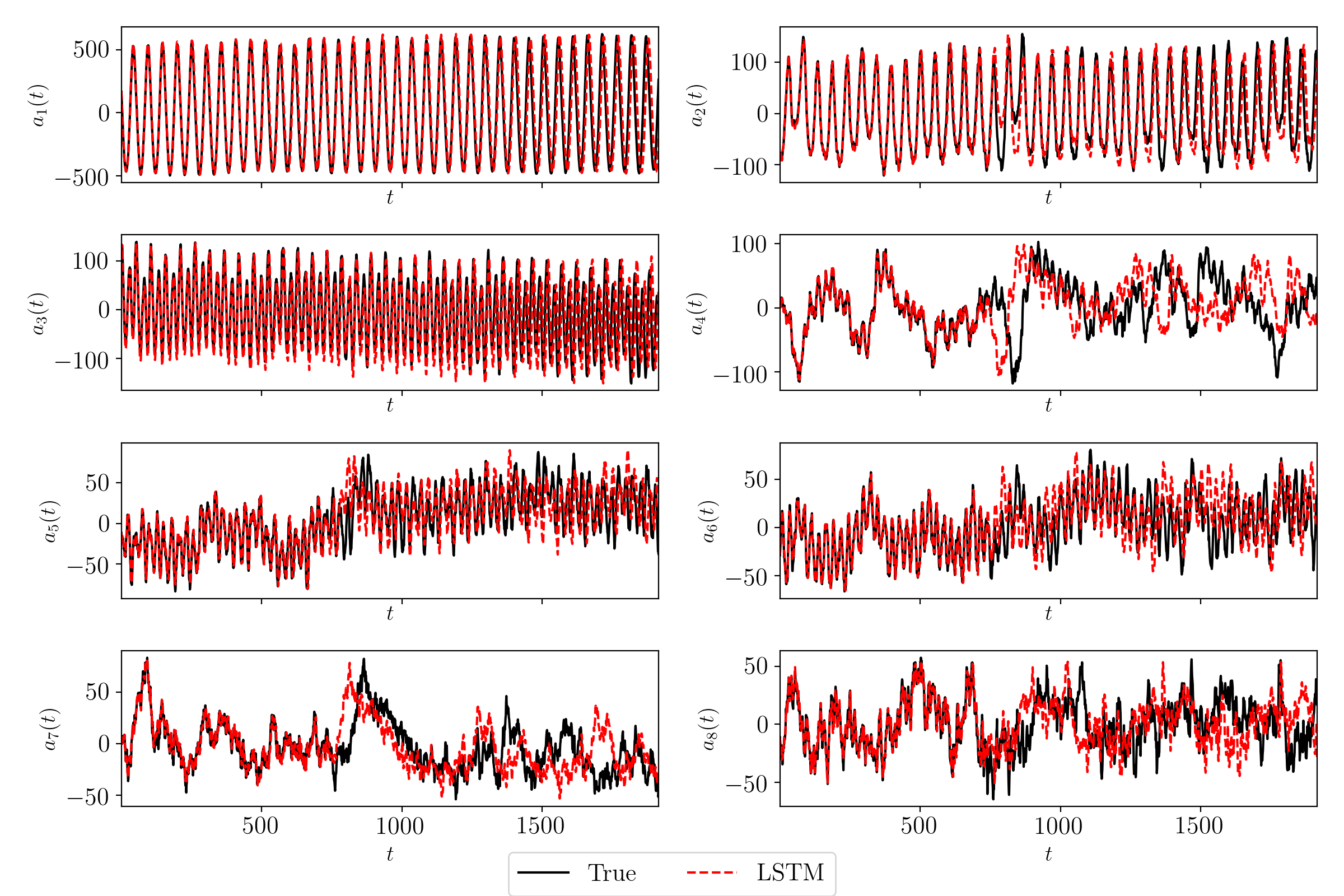}
\caption{Time series prediction of the modal coefficients with the best LSTM network discovered by the proposed GA algorithm.}
\label{fig:time_series}
\end{figure}

The optimal neural network architecture discovered by the genetic algorithm is shown in Figure~\ref{fig:opt_architecture}. Once the best architecture and other hyperparameters are discovered using the GA, the best LSTM network is trained for 1200 epochs with a batch size of 64. During the deployment of the trained network, the auto-regressive method is used to forecast the modal coefficients. The initial condition for modal coefficients for the first 8 time steps (equal to the lookback of the LSTM) is provided. This information is used to predict the modal coefficients at the 9th time step. Then the modal coefficients from the 2nd time step to the 9th time step are used to predict the modal coefficients at the 10th time step. This procedure is repeated up to the final time step, i.e., the 1,914th-time step. Since we use the auto-regressive deployment of the trained LSTM network, only the initial condition corresponding to the LSTM lookback is required. After a few time steps, only the prediction of the LSTM is used to forecast the modal coefficients in the future. In Figure~\ref{fig:time_series}, the true modal coefficients and the predicted modal coefficients are shown. We can see that there is a very good agreement between the true and predicted modal coefficients, especially for the first few modes which are responsible for capturing large-scale fluctuations and seasonal patterns. 

\begin{figure}[htbp]
\centering
\includegraphics[width=0.9\textwidth]{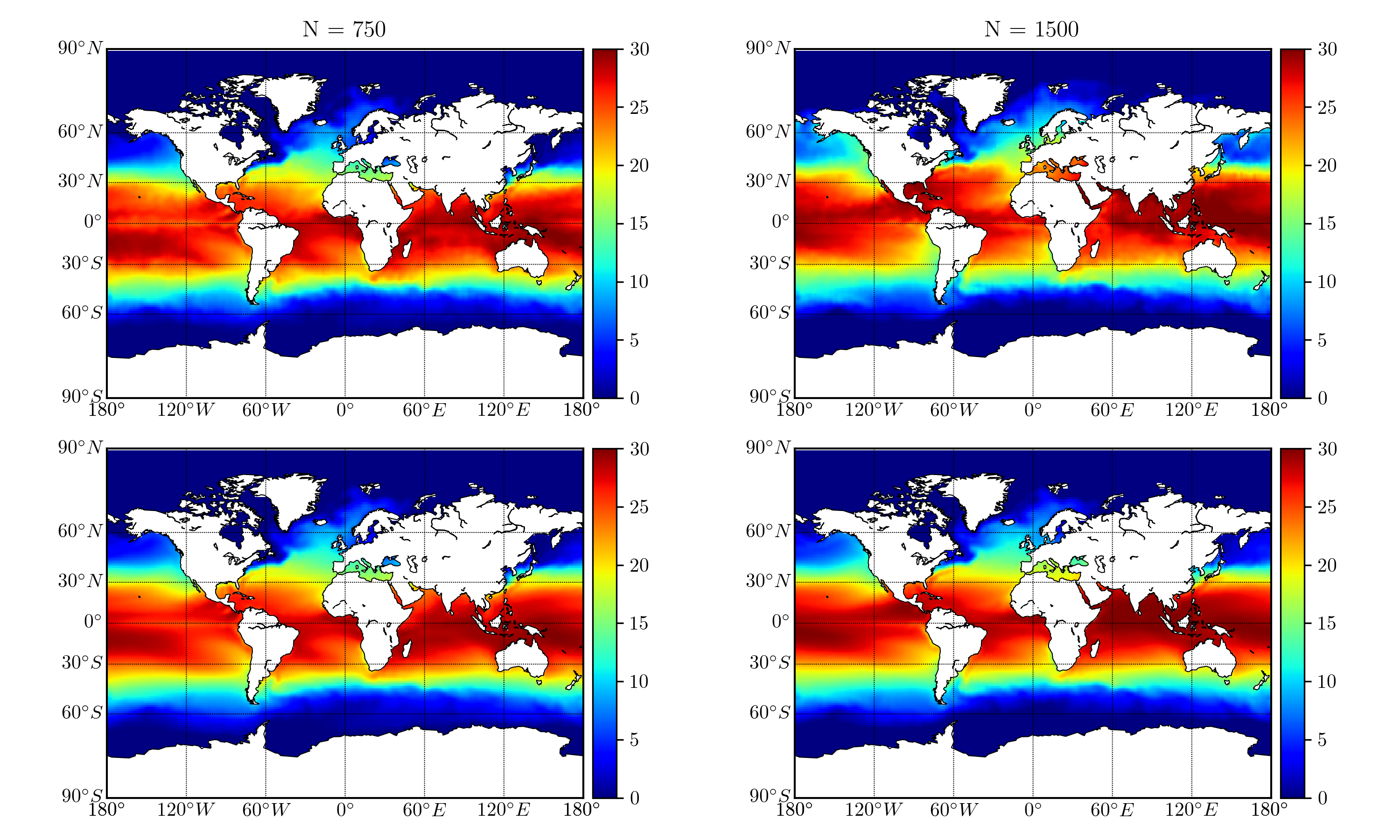}
\caption{Sample averaged temperature field in degrees Celsius at time $T=750$ (left)  and $T=1500$ (right). Top row represent the true temperature field and the bottom row represent the reconstructed temperature field using the ROM.}
\label{fig:fied}
\end{figure}

\begin{figure}[htbp]
\centering
\includegraphics[width=0.9\textwidth]{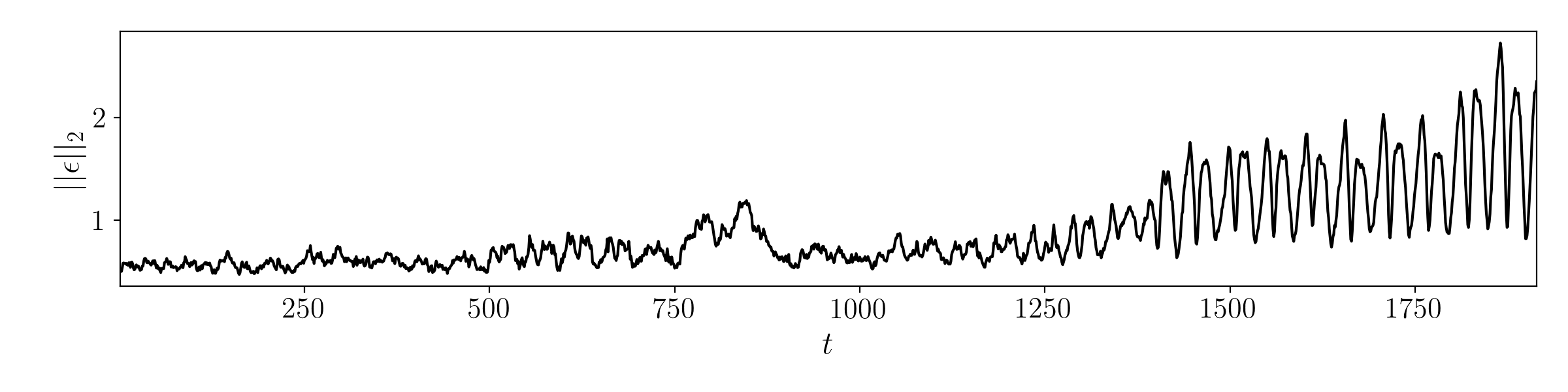}
\caption{Temporal variation of the $L_2$-norm of the difference between the true and predicted temperature field.}
\label{fig:reconstruction}
\end{figure}

Figure~\ref{fig:fied} depicts the true and reconstructed temperature field at two different times. The temperature field is reconstructed using Equation~\ref{eq:rom_construction}, where the average temperature field is computed using the first 1,500 snapshots. The large patterns in the temperature field are accurately captured with the LSTM based ROM which demonstrates the potential of data-driven ROMs for geophysical flows. In Figure~\ref{fig:reconstruction}, the $L_2$-norm of the difference between the true and predicted temperature field is shown. We can observe that the error grows toward the final time, which could be due to the inaccurate prediction of modal coefficients and the inability of the POD basis functions to capture the spatial pattern with high accuracy in the extrapolated time region (i.e., beyond 1500). One of the remedies to this behavior with the LSTM error piling up is to train and deploy the network in a non-auto-regressive manner \cite{maulik2020non}.   

\section{Concluding Remarks} \label{sec:conclusion}
We presented a genetic algorithm (GA) for automatic architecture search and hyperparameter optimization of the long short-term memory (LSTM) neural network for the task of surrogate modeling of geophysical flows. We have successfully demonstrated the surrogate model built using the optimized LSTM is able to forecast the sea-surface temperature (SST) field. To facilitate the training of deeper neural networks, we adopt an encoding strategy in which the LSTM network is designed using small building blocks that incorporate skip connections. The optimized LSTM network can predict the seasonal variation in the SST field accurately without any instability issue for a sufficiently long period. 

We observe that the discrepancy between the true and predicted modal coefficients is higher for the forecast period compared to the training zone. This issue can be addressed either using the non-auto-regressive deployment or transfer learning approach where the LSTM is retrained as the new data becomes available. In the present study, we assume that the number of LSTM cells in each hidden layer is constant. This constraint can be removed and the LSTM network can be optimized with the variable-length encoding strategy for each individual in the population. The variable-length encoding strategy will let us design the neural network architecture that employs a different number of hidden units in each of the LSTM cells and we plan to investigate this in our future studies. Another challenge with the neural architecture search is the computational resources required to train the neural network, and this challenge can be addressed using asynchronous computation.


\section*{Acknowledgments}
This material is based upon work supported by the U.S. Department of Energy, Office of Science, Office of Advanced Scientific Computing Research under Award Number DE-SC0019290. O.S. gratefully acknowledges their support. 
Disclaimer: This report was prepared as an account of work sponsored by an agency of the United States Government. Neither the United States Government nor any agency thereof, nor any of their employees, makes any warranty, express or implied, or assumes any legal liability or responsibility for the accuracy, completeness, or usefulness of any information, apparatus, product, or process disclosed, or represents that its use would not infringe privately owned rights. Reference herein to any specific commercial product, process, or service by trade name, trademark, manufacturer, or otherwise does not necessarily constitute or imply its endorsement, recommendation, or favoring by the United States Government or any agency thereof. The views and opinions of authors expressed herein do not necessarily state or reflect those of the United States Government or any agency thereof.
\bibliography{ref}

\end{document}